\documentclass[aps,prb,reprint,twocolumn,amsmath,amssymb]{revtex4-2}
\usepackage[utf8]{inputenc}
\usepackage{hyperref}
\usepackage{graphics}
\usepackage{graphicx}
\usepackage{braket}
\usepackage{xcolor}
\usepackage{multirow}
\renewcommand{\bf}[1]{\mathbf{#1}}
\renewcommand{\rm}[1]{\mathrm{#1}}
\newcommand{\ketbra}[2]{\ket{#1}\bra{#2}}
\newcommand{\wem}{w_{\rm{em}}}

\begin{document}

\title{Thermal Purcell effect and cavity-induced renormalization of dissipations}
\author{Giuliano Chiriacò}
\affiliation{Dipartimento di Fisica e Astronomia ``Ettore Majorana'', Universit\`{a} di Catania, Catania, Italy}
\email{giuliano.chiriaco@dfa.unict.it}

\date{\today}

\begin{abstract}
In recent years there has been great interest towards optical cavities as a tool to manipulate the properties and phases of embedded quantum materials. Due to the \textit{Purcell effect}, a cavity changes the photon phase space and thus the rate of electromagnetic transitions within the material, modifying the exchange rate of heat radiation with the photon environment. This work derives a simple expression for the radiative heat absorbed by the material, investigates how it changes in the presence of a cavity and shows that it is enhanced dramatically for appropriate cavity geometries. This effect is compared with typical energy dissipation processes, providing a criterion to estimate its impact on the temperature of a material in the cavity and applying it to 1T-TaS$_2$.
\end{abstract}

\maketitle

\emph{Introduction} -- Light-matter interactions in many-body systems have recently sparked great interest as a way to  engineer desirable properties in quantum systems ~\cite{Rev:Schlawin2022,Rev:Garcia-Vidal2021,Mivehvar2021,BlochReview,New:Kurizki2015}. For example, one possibility is the use of photoexcitation to access driven or transient phases not available in equilibrium.\cite{Fausti11,Morrison14,Mitrano16,Kogar20,Nova19,Disa20,Chiriaco18,Chiriaco2020:NAC,Sentef17,Sun20}

Another promising avenue consists in embedding a quantum material inside an optical cavity to modify its equilibrium properties without the use of external illumination. In fact, vacuum effects produced by the cavity modify the electromagnetic environment and mediate interactions in the embedded system. This opens up many fascinating possibilities for the study of new physics in platforms based on solid state, cold atoms or quantum optics systems \cite{New:Kiffner2019c,New:Ashida2020,SC:Chakraborty2021,Ritsch13, Dicke:Landig2016, Cosme18, Georges18}. 

For example, recent experimental advances in molecules \cite{Mol:Zhong2017a,Mol:Feist2018,Mol:Flick2018,Mol:Ribeiro2018,Mol:Rozenman2018b,Mol:Kena-Cohen2019a,Mol:Takahashi2020b,Mol:Wellnitz2020a} and semiconductors \cite{Exc:Feist2015,Exc:Orgiu2015} showed that coupling to a cavity may induce substantial changes in reaction rates \cite{Vibr:Thomas2019c,Lather2019:Vibr,debernardis2023:relaxation}, transport \cite{MB:Hagenmuller2017b,MB:Du2018b,MB:Hagenmuller2018b} and topological properties \cite{Faist19,Faist22}. Meanwhile, theoretical proposals investigated the possibility of using light-matter coupling to control electronic phases, such as superconductivity and ferro-electricity~\cite{Sentef_sciadv2018,Curtis_prl2019,Schlawin_prl2019,Latini_pnas2021,New:Ashida2020}, topological or magnetic phases and quantum spin liquids \cite{Mivehvar2017,Topo:Dmytruk2022, Topo:Appugliese2022,Bacciconi2023:MagnLadder,New:Chiocchetta2021,mercurio2023photon}, and to exploit the entangled light-matter states \cite{ENT:Mendez2020TopologyCavity,Chiriaco2022:Entanglement}.

Among the many consequences of embedding a matter system in a cavity, the \emph{Purcell effect} \cite{Purcell1946} is one of the most intriguing phenomena. It consists in an enhancement of the spontaneous decay rate of atomic transitions when placed inside a resonant cavity, due to the increased photon density of states (phDoS) compared to free space.

The Purcell effect has vast consequences on many aspects of light-matter systems. In particular, since the change in phDoS affects all quantum transitions involving the exchange of photons, it also modifies the rate at which a system exchanges radiative heat with the environment, producing a \emph{thermal Purcell effect}. These effects produce either an enhancement or a suppression of the radiation rate, depending on the cavity phDoS compared to free space. Radiation flows both ways: thermal photons in the environment heat the system up and the system radiates away energy. In this sense, the cavity can act as a shield, that protects the system from heat exchange with the photon environment, or as a magnifying glass that accelerates the rate at which energy is radiated.

A recent remarkable experiment \cite{Fausti_TaS2} showed that coupling to a cavity affects the critical temperature of the metal-insulator transition in the transition metal dichalcogenade 1T-TaS$_2$ \cite{Wilson1975:1TTaS2or,Sipos2008:1TTaS2,Dean2011:Polaronic1T,Wang2020:1TTaS2}. In order to explain the shift in critical temperature, the authors proposed that the Purcell effect results in an effective system temperature different from that of the connected cryostat.

Thus Purcell thermal effects are a crucial factor to consider when investigating the interplay between electronic phases and cavity. Particularly when a phase transition is thermally driven, determining the true system temperature becomes essential. It is then of great importance to formulate a precise theory of how the cavity affects the radiative energy exchange and compare it with other dissipation processes.

\begin{figure}
    \centering
    \includegraphics[width=0.9\columnwidth]{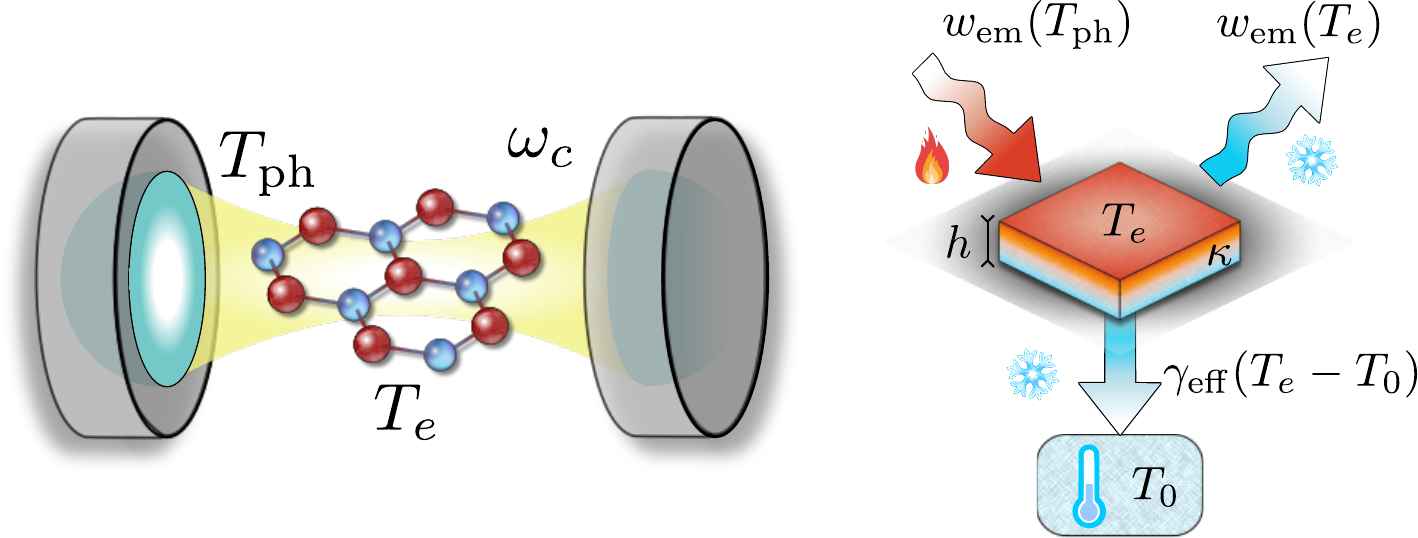}
\put(-220,75){(a)} 
\put(-100,75){(b)}  
    \caption{(a) Sketch of the system: a generic quantum material at temperature $T_e$ is embedded in a cavity with fundamental frequency $\omega_c$; the surrounding photons are at temperature $T_{\rm{ph}}$. (b) Sketch of the radiative energy exchange between the system, the photons and the cryostat. }
    \label{fig:F1CavitySketch}
\end{figure}

Inspired by these considerations, the purpose of this work is to provide a straightforward criterion to evaluate the importance of Purcell thermal effects in cavity experiments. A formula is derived for the rate of energy absorbed by a system via radiative exchange with a surrounding photon bath. The absorbed power is expressed in terms of either known or experimentally accessible quantities, and is then calculated for a system in free space or within a cavity, considering two relevant geometries. Radiative heat exchange and heat dissipation into a cryostat are then compared, demonstrating the substantial impact of thermal effects on the system temperature. Notably, numerical estimates are presented for the case of 1T-TaS$_2$, indicating that thermal effects are crucially dependent on cavity geometry and material properties.

\emph{Radiative heat exchange} -- Let us consider a generic system at temperature $T_e$, in contact with a cryostat at temperature $T_0$; the two temperatures may be different. The system interacts with the electric field of the photon environment through a dipole interaction $H_I=-\bf{\hat d}\cdot\bf{\hat E}$, with $\bf{\hat E}$ the electric field and $\bf{\hat d}$ the dipole operator.

The power per unit of volume absorbed by the system is calculated with the Fermi golden rule \cite{Grosso:Parravicini}
\begin{equation}\label{Eq:AbsHeat}
w=\int d\omega D_{\rm{ph}}(\omega)\frac{\sigma(\omega,T_e)}{\epsilon_0}\omega[n_{\rm{ph}}(\omega)-n_B(\omega,T_e)],
\end{equation}
where $D_{\rm{ph}}(\omega)$ is the phDoS at energy $\omega$, $\sigma(\omega,T_e)$ is the ac electrical conductivity of the system at energy $\omega$, $\epsilon_0$ is the vacuum dielectric permittivity, $n_{\rm{ph}}$ is the photon distribution and $n_B(\omega,T_e)=(e^{\frac{\omega}{k_BT_e}}-1)^{-1}$ is the Bose-Einstein distribution at temperature $T_e$. 

A detailed derivation of Eq. \eqref{Eq:AbsHeat} is given in Appendix \ref{App:RadHeat}, but it can be interpreted as follows: $n_{\rm{ph}}$ ($n_B$) are the number of electromagnetic fluctuations carrying energy into (out of) the system, $\omega$ is their energy and $\epsilon_0/\sigma$ is the typical lifetime of such excitations \footnote{Indeed this is exactly the lifetime of a charge fluctuation in a metal, as derived from the Maxwell equations.}. Thus summing them over the photon modes yields the power density absorbed by the material due to exchange of electromagnetic radiation \footnote{While Eq. \eqref{Eq:AbsHeat} bears many resemblances with the energy density of blackbody radiation, it should rather be thought of as the power dissipated by Joule heating due to the electric field of thermal photons.}. When $w$ is positive the system absorbs energy, while for $w<0$ it cedes energy to the environment.

Equation \eqref{Eq:AbsHeat} is valid for a generic (even non-equilibrium) photon distribution; if the photons are in equilibrium at temperature $T_{\rm{ph}}$, then $n_{\rm{ph}}=n_B(\omega,T_{\rm{ph}})$, $w$ vanishes for $T_{\rm{ph}}=T_e$ and $w>0$ for $T_{\rm{ph}}>T_e$ as it should. In the following, the photon environment is assumed to be thermal at temperature $T_{\rm{ph}}$.

The remarkable feature of Eq. \eqref{Eq:AbsHeat} is that all microscopic details about the nature of the system are encoded inside $\sigma$, which is directly measurable through spectroscopic or transport measurements. Similarly all details about the cavity geometry and parameters are captured by $D_{\rm{ph}}$, which can be calculated theoretically. The distribution of the photon bath is determined by the environment temperature. The system temperature $T_e$ can be inferred via the energy tail of ARPES spectra or via infrared imaging.

In the case of a temperature independent conductivity, the absorbed power is simply the difference $w=\wem(T_{\rm{ph}})-\wem(T_e)$ between two emitted powers 
\begin{equation}
\label{Eq:wem}
\wem(T)=\int d\omega D_{\rm{ph}}(\omega)\frac{\sigma(\omega)}{\epsilon_0}\omega n_B(\omega,T),
\end{equation}
so that any analysis of $w$ reduces to analyzing $\wem$.

\emph{Analysis of Purcell thermal effects} -- The interplay between the phDoS and the conductivity $\sigma$ is crucial in determining the absorbed power and the magnitude of the thermal Purcell effect. In the following, the conductivity is assumed constant in frequency $\sigma(\omega)=\sigma_0$; this model is able to capture most of the qualitative physics at play. More sophisticated models considered in Appendix \ref{App:Condu} exhibit a similar phenomenology.

The phDoS is $D_{\rm{ph}}(\omega)=\sum_n\frac{1}{V_n}\delta(\omega-\omega_n)$ -- with $\omega_n$ the energy of cavity mode $n$ and $V_n$ the \emph{effective} volume occupied by the mode \footnote{In most cases $V_n$ is the same for each mode $n$. However, for confined cavities it depends on $n$.}, which can be smaller than the physical volume of the cavity. Three paradigmatic cases are considered:

\emph{(a)} photons in free space. The phDoS is the usual one calculated for black-body radiation $D^{(a)}_{\rm{ph}}(\omega)=\frac{\omega^2}{\pi^2\hbar^3c^3}$.

\emph{(b)} A Fabry-Perot cavity with infinite planar mirrors, fundamental energy $\omega_c$ and high quality factor. The phDoS is $D_{\rm{ph}}^{(b)}(\omega)=\frac{\omega\omega_c}{\pi^2\hbar^3c^3}(\lfloor\omega/\omega_c\rfloor+1/2)$, with $\lfloor\rfloor$ the integer part. 

\emph{(c)} A confined cavity where spherical mirrors squeeze the modes in the radial direction. The phDoS is $D_{\rm{ph}}^{(c)}(\omega)=\frac{\omega_{\rho}^2\omega_c}{\pi^2\hbar^3c^3}\sum_{q,l,m}\delta(\omega-\omega_c(1+q+l+m))$, where $q,l,m$ are the longitudinal and radial (integer) quantum numbers, $\omega_c$ is the fundamental energy and $\omega_{\rho}$ is an energy scale associated to the transverse radial size of the photon modes \cite{Photo:book,Papageorge2016:ConcHighTM,Utama2021:ConcHighTM}. This phenomenology also applies to cavities where the radial confinement is due to other factors, such as objects altering the fields inside the cavity (e.g. the system itself, the sample holder, the cryostat); the derivation of the phDoS is detailed in Appendix \ref{App:phDoS}.

\begin{table}[b]
\centering
\begin{tabular}{c c c c c} \hline
\rule{0pt}{0.3cm}\qquad & $D_{\rm{ph}}^{(a)}$ & $D_{\rm{ph}}^{(b)}$ & $D_{\rm{ph}}^{(c)}$ &\\
\hline
\qquad & \multirow{2}{4em}{\centering$\sim T^4$}  & \quad$\sim \wem^{I,0}$ & \quad$\sim\omega_{\rho}^2T^4/\omega_c^2$ &\quad$\omega_c\ll T$\quad\\ 
\qquad & & \quad$\sim\omega_c T^3$ & \quad$\sim\omega_{\rho}^2\omega_c^2e^{-\omega_c/k_BT}$ & \quad$\omega_c\gg T$\quad\\
\hline
\end{tabular}
    \caption{Asymptotic behaviors of $\wem(T)$ for $\omega_c\ll T$ and for $\omega_c\gg T$.}
    \label{tab:Tab1}
\end{table}

\begin{figure}[!t]
    \centering
     \hspace*{-0.045\columnwidth} 
    \includegraphics[width=1.08\columnwidth]{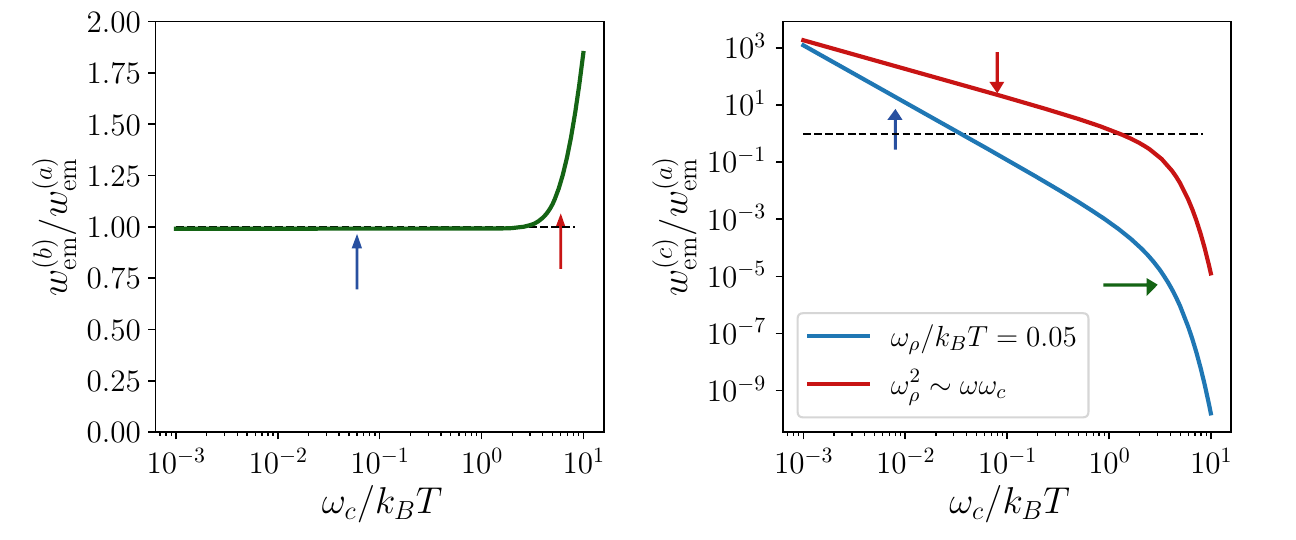}
    \put(-264,96){(a)} 
\put(-134,96){(b)} 
\put(-200,39){$\sim1$} 
\put(-175,46){$\sim\omega_c/T$} 
\put(-100,69){$\sim\omega_{\rho}^2/\omega_c^2$} 
\put(-55,95){$\sim1/\omega_c$} 
\put(-69,50){$\sim e^{-\frac{\omega_c}T}$} 
\caption{Enhancement of the radiative power $\wem$ compared to free space (grey dashed line) as function of $\omega_c/T$ for (a) planar cavity and (b) confined cavity for $\omega_{\rho}/k_BT=0.05$ (blue) and a $\omega_c$-dependent model for $\omega_{\rho}$ (red). Inset text shows the scaling behaviors at $\omega_c\ll k_BT$ and $\omega_c\gg k_BT$.}
    \label{fig:F2wemPlot}
\end{figure}
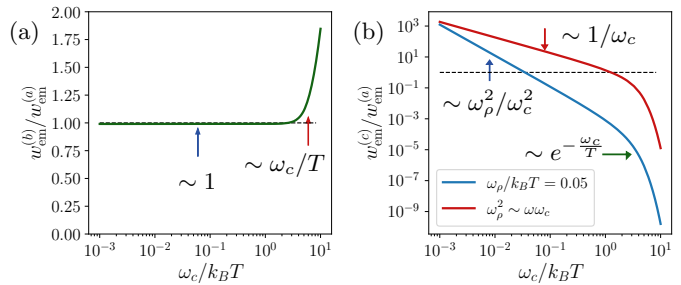

Since $\omega_c\lfloor\omega/\omega_c\rfloor\sim\omega$, $D_{\rm{ph}}^{(b)}$ is the same order of magnitude of $D_{\rm{ph}}^{(a)}$ and $D_{\rm{ph}}^{(b)}\rightarrow D_{\rm{ph}}^{(a)}$ when $\omega_c\rightarrow0$, which corresponds to opening the cavity mirrors. In the limit $\omega_c\gg T$, the small $\omega$ behavior $D_{\rm{ph}}^{(b)}\sim\omega\omega_c$ (due to the transverse modes \cite{Dutra1996:Purcell,pannirsivajothi2024:blackbody,fassioli2024:controlling}) becomes important and leads to $\wem^{(b)}\sim\omega_cT^3$; the asymptotic regimes are summarized in Tab. \eqref{tab:Tab1}. The ratio $\wem^{(b)}/\wem^{(a)}$, plotted in Fig. \ref{fig:F2wemPlot}a), is $\sim1$ for small $\omega_c$ and linear $\sim\omega_c/T$ at $\omega_c\gtrsim T$.

The magnitude of the phDoS in the \emph{(c)} case crucially depends on $\omega_{\rho}$, connected to the typical radial size $r_0$ of the photon modes by $\omega_{\rho}=2\hbar c/r_0$, see Appendix \ref{App:phDoSconc}. For nearly concentric cavities the modes are extremely confined, and $r_0\rightarrow0\Rightarrow\omega_{\rho}\rightarrow\infty$; for millimetric cavities, $\omega_{\rho}$ can easily assume values in the THz range. The ratio $\omega_{\rho}/\omega_c$ characterizes the cavity geometry -- more precisely the squeezing of the modes in the radial vs longitudinal direction. Depending on this ratio, $D^{(c)}_{\rm{ph}}$ and $\wem^{(c)}$ may be greatly enhanced compared to free space.

The asymptotic behaviors are investigated analytically. For large $\omega_c/T$, all terms in the sum are suppressed exponentially and the fundamental mode ($n=l=m=0$) is the leading term: $\wem^{(c)}\sim\omega_{\rho}^2\omega_c^2e^{-\omega_c/T}$. For small $\omega_c/T$, the sum in the phDoS is approximated as an integral, yielding $D^{(c)}_{\rm{ph}}\sim\omega_{\rho}^2\omega^2/\omega_c^2$; upon integrating $\int d\omega \omega^3n_B(\omega)\sim T^4$, one obtains $\wem^{(c)}\sim\omega_{\rho}^2T^4/\omega_c^2$, which is renormalized by a factor $\omega_{\rho}^2/\omega_c^2$ compared to $\wem^{(a)}$.

The striking $\omega_{\rho}^2/\omega_c^2$ factor causes large thermal effects at small $\omega_c$, i.e. bigger cavities. Intuitively, this arises from the confinement in the two orthogonal directions, which contribute $\sim1/\omega_{\rho}^2$ to the modes volume but still account for a number $\sim1/\omega_c^2$ of modes.

In practice, there are some limitations. Firstly, a concentric cavity becomes unstable when the mirrors are placed too far apart, limiting how small $\omega_c$ can be \cite{Photo:book,Nguyen2018:ConcCavLastRes}. Moreover, in many cases changing $\omega_c$ affects the cavity geometry and the radial size of the modes, i.e. $\omega_{\rho}$ depends on $\omega_c$ due to the geometry of the cavity. As showed in Appendix \ref{App:phDoS}, it is expressed $\omega_{\rho}^2=s^2\omega\omega_c$ -- where $s$ is a cavity geometric factor -- meaning that the asymptotic limits of $\wem^{(c)}$ become $\sim T^5/\omega_c$ and $\omega_c^4e^{-\omega_c/k_BT}$. For completeness, both cases of constant $\omega_{\rho}$ and $\omega_c$-dependent $\omega_{\rho}$ are considered; however, I will use the parameter $\omega_{\rho}$ as an indicator of the radial confinement of the modes.

The behavior of $\wem^{(c)}$ is plotted in Fig. \ref{fig:F2wemPlot}b). In both cases, the value of $\wem$ shows a power-law behavior in $\omega_c$ -- with an enhancement compared to free space at low $\omega_c$ -- and an exponential decay at large $\omega_c$.

\emph{Comparison with heat dissipation} -- The absorbed power in Eq. \eqref{Eq:AbsHeat} should be compared with the other relevant energy dissipation processes.

The only dissipation process is thermal conduction into the cryostat, since convective energy transfer can usually be neglected. Normally heat conduction is a very efficient dissipative process and dominates compared to radiative heating, but its magnitude depends on the contact surface between the system and the cryogenic reservoir. Since radiative heating mostly occurs throughout the entire volume of the system \footnote{In principle, radiative heating occurs near the surface of the system because of the skin effect. However, for typical semimetals and metals and frequencies in the GHz-THz range, the skin depth is of the order of $\rm{\mu m}$, i.e. comparable with the thickness of most samples, so that radiative heating can be approximated as a volume process.}, the thickness of the sample is crucial in determining the dominant process.

To be concrete, the system is at temperature $T_e<T_{\rm{ph}}$, has thickness $h$, thermal conductivity $\kappa$ and is in contact with a reservoir at temperature $T_0<T_e$, see Fig. \ref{fig:F1CavitySketch}b). The thermal current density flowing from the system to the reservoir is $\sim\kappa(T_e-T_0)/h$; this dissipation is roughly equivalent to a volume process with effective rate $\gamma_{\rm{eff}}\sim\kappa/h^2$ -- i.e. dissipation is less efficient for thicker systems \cite{Chiriaco2020:PolarHeating}. Balancing the thermal conduction out of the system, with the absorbed radiative power, leads to the energy balance equation:
\begin{equation}
\label{Eq:HeatBalance}
\gamma_{\rm{eff}}(T_e-T_0)=w=\wem(T_{\rm{ph}})-\wem(T_e).
\end{equation}

In an experiment, one observes the temperature $T_0$ which is in general different from $T_e$; this difference is small when $w/\gamma_{\rm{eff}}\ll T_0$ or roughly $\gamma_{\rm{eff}}\gg \wem(T_{\rm{ph}})/T_{\rm{ph}}$. For free space, $\wem$ is calculated exactly $\wem(T)=\frac{\sigma_0}{\epsilon_0}\frac{4\sigma_{SB}}cT^4$ -- where $\sigma_{SB}$ is the Stefan-Boltzmann constant. Thermal conduction is then dominant compared to radiative heating when $\eta_{/c}\equiv\frac{\wem^{(a/c)}(T_{\rm{ph}})}{\gamma_{\rm{eff}}T_{\rm{ph}}}\ll1$, i.e. for free space and confined cavities respectively:
\begin{equation}
\label{Eq:CriterionFreeSpace}
\eta_a=\frac{\sigma_0}{\epsilon_0}\frac{4}c\frac{\sigma_{SB}T_{\rm{ph}}^3}{\kappa/h^2}\ll1; \qquad
\eta_c\equiv\eta_a\frac{\omega_{\rho}^2}{\omega_c^2}\ll1.
\end{equation}
When both criteria in Eq. \eqref{Eq:CriterionFreeSpace} are satisfied, the thermal Purcell effect is negligible at the practical level.

Now consider a system undergoing a phase transition at temperature $T_e=T_c$, a case relevant for Ref. \cite{Fausti_TaS2}. If $T_c$ is not affected by the cavity, the transition is observed at the temperature of the cryostat \cite{Chiriaco2020:PolarHeating,Chiriaco2018:CDW}:
\begin{equation}
\label{Eq:TcObs}
T_c^{obs}=T_0=T_c-\frac{\wem(T_{\rm{ph}})-\wem(T_c)}{\gamma_{\rm{eff}}},
\end{equation}
so that $T_c^{obs}\leq T_c$ and $T_c^{obs}\rightarrow T_c$ when the criteria of Eq. \eqref{Eq:CriterionFreeSpace} are satisfied. Since $\wem^{(b)}/\wem^{(a)}$ increases with $\omega_c$ it leads to a smaller $T_c^{obs}$ at large $\omega_c$, the opposite trend of the experimental data in \cite{Fausti_TaS2} which show an increase of $T_c^{obs}$ with $\omega_c$. On the other hand, $\wem^{(b)}/\wem^{(a)}$ produces a trend of $T_c^{obs}$ in agreement with the data, so that only confined cavities are considered from now on.

\begin{figure}[t]
    \centering
    \includegraphics[width=1.03\columnwidth]{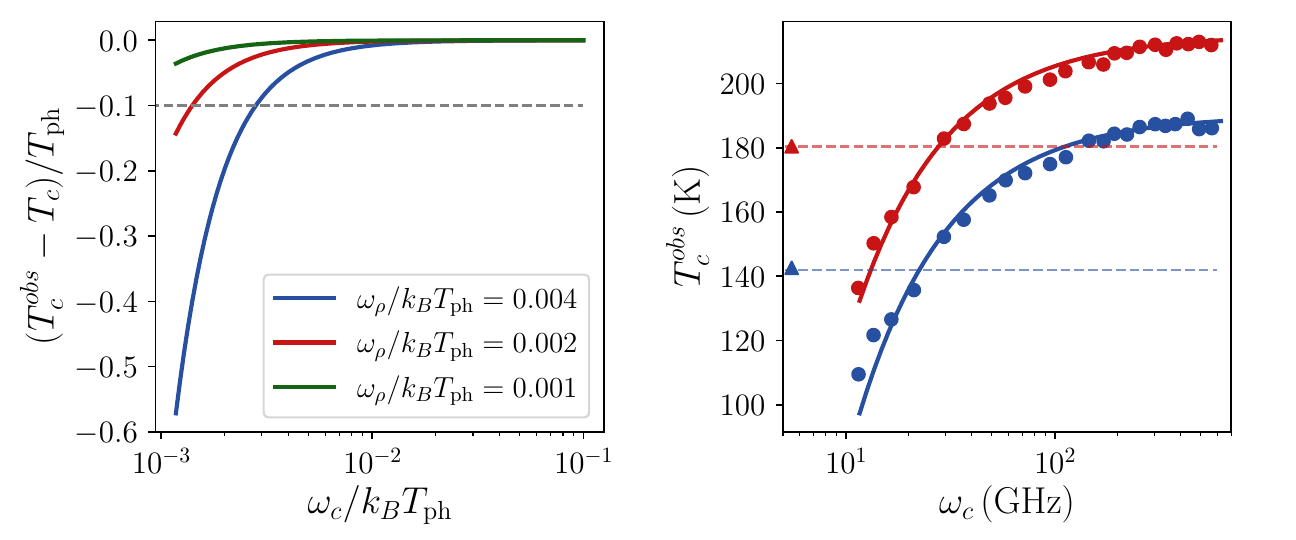}
\put(-250,96){(a)} 
\put(-125,96){(b)}  
\caption{(a) Relative difference in the observed critical temperature as function of $\omega_c/T_{\rm{ph}}$ at various values of $\omega_{\rho}$, for $\eta_a=0.1$ and $T_c=0.5T_{\rm{ph}}$. The grey dashed line is $T_c^{obs}$ in free space. (b) Theoretical fit (solid lines) of the experimental data (circles) of the observed critical temperature in 1T-TaS$_2$, as reproduced from \cite{Fausti_TaS2}; red (blue) refers to the heating (cooling) hysteresis data. The triangles and dashed lines indicate the experimental and theoretical value of $T_c^{obs}$ in free space, respectively. The fit is performed using a $\omega_c$-dependent model for $\omega_{\rho}$ with parameters $\eta_a^{(h)}=0.115$ ($\eta_a^{(c)}=0.16$) and $s^{(h)}=0.049$ ($s^{(c)}=0.044$).}
    \label{fig:F3HeatBalance}
\end{figure}

In Fig. \ref{fig:F3HeatBalance}a), $T_c^{obs}$ is plotted as function of $\omega_c/T_{\rm{ph}}$ for a confined cavity at different values of constant $\omega_{\rho}$ and $\eta_a=0.1$, showing a trend compatible with that observed in \cite{Fausti_TaS2}. A quantitative fit of the experimental data from 1T-TaS$_2$ can be performed employing both models for $\omega_{\rho}$, and finding that a $\omega_c$-dependent $\omega_{\rho}$ model better fits the data. The fitting parameters are then $\eta_a$ and $s$, and the results of the fit are plotted in Fig. \ref{fig:F3HeatBalance}b). The model fits very well the cooling and heating branches of the hysteresis curve.

The value of $\eta_a$ could also be estimated for 1T-TaS$_2$ using the system parameters. The dc conductivity of the metallic phase is $\sigma_0\approx10^5\,\rm{\Omega}^{-1}\rm{m}^{-1}$ \cite{Disalvo1977:Electro1T,Martino2020:Electro1T} and $T_{\rm{ph}}=300\,\rm{K}$, yielding $\wem^{(a)}\approx7\cdot10^{10}\,\rm{W/m^3}$. The thermal conductivity is $\kappa\approx2\,\rm{W}\rm{m^{-1}K^{-1}}$ \cite{Nunez1985:Thermo1T,Liu2020:Thermo1T} and the thickness is $h\sim10\,\rm{\mu m}$ leading to inefficient dissipation $\gamma_{\rm{eff}}\sim2\cdot10^{10}\,\rm{W}\rm{m^{-3}K^{-1}}$. This yields $\eta_a=w_{\rm{em}}^{(a)}/(\gamma_{\rm{eff}}T_{\rm{ph}})\approx0.011$ -- and order of magnitude smaller than the fit; this value of $\eta_a$ produces a small downshift of $T_c^{obs}$ in free space, $\Delta T_c=T_c^{obs}-T_c\approx-3\,\rm{K}$, ten times smaller than the observed shift. This discrepancy is likely due to an overestimation of $\gamma_{\rm{eff}}$, which in \cite{Fausti_TaS2} is further decreased by the silicon nitride (SiN) membrane that physically connects the sample to the cryostat. An accurate estimate of $\gamma_{\rm{eff}}$ from experimental parameters require knowing precisely the SiN conductivity and thickness; an analysis along these lines is performed in \cite{Fausti_TaS2}.

The value of $s$ obtained from the fit corresponds, for typical energies $\omega\sim k_BT_{\rm{ph}}\sim6$ THz and $\omega_c\sim20$ GHz, to $\omega_{\rho}=s\sqrt{\omega\omega_c}\sim20$ GHz, i.e. a typical radial size $r_0\sim5$ mm compatible with a light confinement of the modes in the cavity of Ref. \cite{Fausti_TaS2}.

\emph{Discussion} -- These results show that the radial confinement of the cavity photons (relatively to longitudinal confinement) plays a crucial role in the thermal Purcell effect and can result in a substantial renormalization of radiative dissipation. Even modest values of $\omega_{\rho}$, corresponding to a rather shallow radial confinement of the modes, can provide large effects on the renormalization of the system temperature. In this sense, the cavity acts like a magnifying glass, focusing hot photons onto the system and considerably heating it up.

Both the qualitative behavior of $T_c^{obs}$ and a quantitative fit of the data, indicate a major role of the thermal Purcell effect in the experimental observations on 1T-TaS$_2$. It appears that the low thermal conductivity and the thickness of the sample and membrane make 1T-TaS$_2$ very sensitive to even small confinement effects.

In contrast to \cite{Fausti_TaS2} the present model predicts that planar cavities produce a trend of $T_c^{obs}$ vs $\omega_c$ opposite to the data. The discrepancy originates from the phDoS, which should include the contribution from the transverse modes. These modes are invisible to any spectroscopy of the cavity conducted with light incident perpendicularly to the mirrors, which only probes the longitudinal modes; nonetheless, they couple to the electrons and contribute to radiative energy exchange. Detailed analyses of such scenario are presented in \cite{fassioli2024:controlling,pannirsivajothi2024:blackbody}.

Remarkably, large values of $\omega_{\rho}$ increase the effective light-matter coupling, so that strong thermal effects also mean strong renormalizations of the matter free energy -- via the real part of the dielectric constant. As also suggested in \cite{Fausti_TaS2}, a renormalized free energy can shift the transition temperature, an effect that is enhanced for large $\omega_{\rho}$. This scenario is different from ultra-strong coupling to a single mode \cite{FornDiaz2019:RMPultrastrong}, as it involves coupling to many (confined) modes. Indeed it would be surprising that weak interactions with a single mode can modify the thermodynamical properties of the material \cite{Pilar_2020:thermo,andolina2022deep,Saez2023:VacuumShift}.

Given its role in determining the magnitude of the thermal Purcell effect, the value of the photon temperature $T_{\rm{ph}}$ is a delicate issue \cite{Rüting2017:Tph,Portugal2023:Tph}, especially when the cavity mirrors are at a temperature $T_{\rm{mir}}$ different from room temperature $T_{\rm{env}}$ \cite{Fausti_TaS2,Jarc2022:cryogenic}. The cavity is essentially a driven lossy resonator: environment photons at $T_{\rm{env}}$ are pumped in and leak out after a time proportional to the quality factor $Q$; thus photons in cavities with higher $Q$ -- i.e. more reflective mirrors -- are trapped for longer and thermalize closer to $T_{\rm{mir}}$. On the other hand, the reflection of a photon upon a perfect mirror is an elastic collision, which does not contribute to thermalization; imperfect mirrors allow for thermalization, but also decrease $Q$. Thus it seems likely that $T_{\rm{ph}}$ assumes a value between $T_{\rm{mir}}$ and $T_{\rm{env}}$; an open system dynamics treatment should provide a precise answer, but is beyond the scope of this work.

\emph{Conclusions} -- An expression for Purcell thermal effects was derived, quantifying the radiative energy exchanged between a system embedded in a cavity and the photon environment. For a confined cavity, the energy exchange rate is suppressed or enhanced (depending on its geometry) compared to free space. The enhancement is due to the squeezing of the modes volume in the transverse direction and is substantial at low cavity frequencies, but gets exponentially suppressed at high frequency.

This energy exchange process was confronted with thermal conduction into a cryostat, demonstrating that the two effects are comparable for confined cavities and thick samples. In such scenario, dissipation is inefficient and the temperature observed from the cryostat significantly deviates from the actual system temperature. A quantitative criterion is provided to estimate the importance of thermal effects in experimental contexts.

The observations in 1T-TaS$_2$ can be quantitatively explained by a fit based on a confined cavity model,  confirming the impact of the thermal Purcell effect on the system. On the other hand, a planar configuration exhibits a different trend, stressing the importance of accurately characterizing the cavity geometry. The model also predicts a strong dependence on the thickness of the sample and on the temperature of the photon environment, so it would be interesting to perform experiments on thin films or with cavities fully enclosed in a cryostat.

\emph{Acknowledgements} -- I am grateful to Gian Marcello Andolina, Zeno Bacciconi, Daniele De Bernardis and Daniele Guerci for fruitful discussions. I thank Marcello Dalmonte, Francesco Pellegrino and Emily Tiberi for useful comments. This work is supported by ICSC – Centro Nazionale di Ricerca in High-Performance Computing, Big Data and Quantum Computing under project E63C22001000006. I acknowledge the CINECA award under the ISCRA initiative, for the availability of high-performance computing resources and support.

\bibliographystyle{apsrev4-1}
\bibliography{CavityHeating}

\onecolumngrid

\pagebreak[4]

\appendix

\section{Calculation of the radiative heat power}\label{App:RadHeat}

The goal of this section is to calculate the radiative power density exchanged by a generic matter system at temperature $T_e$ with a photon environment described by a density of states (phDoS) $D_{\rm{ph}}(\omega)$ and a distribution function $n_{\rm{ph}}(\omega)$. I follow the treatment of optical transitions in \cite{Grosso:Parravicini} (also outlined in the Supplemental Material of \cite{Fausti_TaS2} to calculate the polarizability of the matter system). The first step is to calculate the transition rates of a generic optical transition and the power absorbed by the system in the presence of an electric field; relating it to the Joule heating due to a classical electric field, it is possible to express the matter contribution in terms of the conductivity. This is then used for the calculation of the power in the presence of a quantized electric field. This allows to make general predictions without having to specify the microscopic details of the system.

Consider a many body systems composed of electrons described by a generic Hamiltonian $H_e$; the generalization to bosons or to a system with both kinds of degrees of freedom is not difficult. The electrons are subject to an electric field $\bf{\hat E}$ -- which can be generated by a classical field or by quantized photons -- and the interaction with the field is written in the dipole approximation. For weak fields the quadratic diamagnetic term can be ignored. The interaction Hamiltonian is %
\begin{equation}\label{Eq:HI}
    H_I=-\bf{\hat d}\cdot\bf{\hat E},
\end{equation}
where $\bf{\hat d}=(\hat d_x,\hat d_y,\hat d_z)$ is the dipole operator of the electrons. The electronic system is assumed to be istotropic and the dipole operator to be along the direction of $\bf E$. The case of anisotropic system is a simple generalization.

I first assume that the electric field is classical, i.e. $\bf E=\text{Re}(\bf{E_0}e^{i\bf q\cdot\bf r-i\omega t})$, and calculate the emission/absorption transition rate from an initial eigenstate $\ket{i}$ of $H_e$ to a final state $\ket{f}$ with energies $\epsilon_{i/f}$. The space dependence of $\bf E$ can be neglected since the typical wavevectors of a photon are much smaller than electronic wavevectors -- i.e. the speed of light is much larger than the Fermi velocity.

Within perturbation theory, the Fermi golden rule yields
\begin{gather}\label{Eq:GemabsClass}
    \Gamma_{em/ab}=\frac{2\pi}{\hbar}\sum_f|\langle i|\hat d|f\rangle|^2E_0^2f(\epsilon_i)(1-f(\epsilon_f))\delta(\epsilon_f-\epsilon_i\pm\hbar\omega),
\end{gather}
where $f$ is the occupation function of the electrons, given in equilibrium by the Fermi-Dirac distribution $f(\epsilon)=(e^{\epsilon/T_e}+1)^{-1}$.

The power absorbed per unit of volume is obtained by summing over all possible initial states the difference between absorption and emission rate times the photon energy $\omega$:
\begin{gather}
\notag    
w=\frac1V\sum_i\omega(\Gamma_{ab}-\Gamma_{em})=\frac1V2\pi\omega\sum_{i,f}|\langle i|\hat d|f\rangle|^2E_0^2f(\epsilon_i)(1-f(\epsilon_f))[\delta(\epsilon_f-\epsilon_i-\hbar\omega)-\delta(\epsilon_f-\epsilon_i+\omega)];\\
\label{Eq:Wclassical}    
w=\frac1V\frac{2\pi\omega}{\hbar}\sum_{i,f}|\langle i|\hat d|f\rangle|^2E_0^2[f(\epsilon_i)-f(\epsilon_f)]\delta(\epsilon_f-\epsilon_i-\omega),
\end{gather}
where $V$ is the system volume and the last line follows by exchanging $f$ with $i$. The absorbed power is essentially the Joule heating, which is connected to the ac conductivity, so that $w=2\sigma E_0^2$ (the factor $2$ comes from the time average of the $\omega$ and $-\omega$ Fourier components). Using $d_{if}^2\equiv|\langle i|\hat d|f\rangle|^2$ one can write
\begin{gather}
\notag    w=E_0^2\frac1V\int d\varepsilon\frac{2\pi\omega}{\hbar}\sum_{i,f}d_{if}^2[f(\varepsilon)-f(\varepsilon+\omega)]\delta(\varepsilon-\epsilon_i)\delta(\epsilon_f-\varepsilon-\omega)=2\sigma(\omega,T_e)E_0^2;\\
\label{Eq:sigma}    \sigma(\omega,T_e)=\frac{\pi\omega}{\hbar}\int d\varepsilon [f(\varepsilon)-f(\varepsilon+\omega)]\frac1V\sum_{i,f}d_{if}^2\delta(\varepsilon-\epsilon_i)\delta(\epsilon_f-\varepsilon-\omega),
\end{gather}
where $\sigma(\omega,T_e)$ is the ac conductivity of the system at energy $\omega$ and temperature $T_e$.

In the case of a quantized electromagnetic field, the photon Hamiltonian and the electric field are given by a sum over the electromagnetic modes $n$
\begin{gather}
\label{Eq:HphEph}
H_{\rm{ph}}=\sum_n\omega_na^{\dagger}_{n}a_{n};\qquad
\bf{\hat E}=i\sum_n\sqrt{\frac{\omega_n}{2\epsilon_0V_n}}(\bf e_n(\bf r)a_{n}e^{-i\omega_n t/\hbar}-\bf e^*_n(\bf r)a^{\dagger}_{n}e^{+i\omega_n t/\hbar}),
\end{gather}
with $\omega_n$ the energy and $a_n$ the annihilation operator for mode $n$.  The polarization vector $\bf e_n$ contains the space profile and is normalized to the mode volume $V_n$: $\int|\bf e_n(\bf r)|^2d\bf r=V_n$. Again the space dependence in $\bf e_n$ can be neglected because it varies over lengthscales much bigger than the dimensions of a unit cell of the electronic system so that $|\bf e_n|^2\approx1$.

Within the FGR formula it is necessary to sum over all initial and final photon states. In the weak coupling limit, the eigenstates of the total Hamiltonian are factorized into electronic states and photon states $\ket{i}=\ket{i}_e\otimes\ket{i}_{\rm{ph}}$, so that $|\langle i|H_I|f\rangle|^2=|\langle i|\bf{\hat d}|f\rangle_e\cdot\langle i|\bf{\hat E}|f\rangle_{\rm{ph}}|^2= d_{if}^2|\langle i|\bf{\hat E}|f\rangle_{\rm{ph}}|^2$  and the sum can be carried out independently over electrons and photons. 

The sum over the initial and final photon states leads to different results for the case of emission and absorption. The $a_{n}$ term is responsible for the absorption of photons, while the  $a_{n}^{\dagger}$ term is responsible for the emission of photons. In the case of initial photon mixed states, one has to weight them with their occupation probability $p_{n,i}$.
For emission processes
\begin{gather}\label{Eq:aadaga}
\sum_{i,f}p_{n,i}|\langle f|a_n^{\dagger}|i\rangle_{\rm{ph}}|^2=\sum_{i}p_{n,i}\bra{i}a_n\sum_f\ketbra{f}{f}_{\rm{ph}}a^{\dagger}_n\ket{i}_{\rm{ph}}=
\sum_{i}p_{n,i}\bra{i}a_na^{\dagger}_n\ket{i}_{\rm{ph}}=\langle a_na^{\dagger}_n\rangle_{\rm{ph}}=1+n_{\rm{ph}}(\omega_n),
\end{gather}

where $\sum_f\ketbra{f}{f}_{\rm{ph}}={1}_{\rm{ph}}$ is the resolution of the identity and $n_{\rm{ph}}$ is the photon number occupation at energy $\omega_n$. Similarly for absorption processes one can write
\begin{gather}\label{Eq:adagaa}
\sum_{i,f}p_{n,i}|\langle f|a_n|i\rangle_{\rm{ph}}|^2=\langle a_n^{\dagger}a_n\rangle_{\rm{ph}}=n_{\rm{ph}}(\omega_n).
\end{gather}
Combining Eqs. \eqref{Eq:aadaga}-\eqref{Eq:adagaa} with Eq. \eqref{Eq:GemabsClass} one can write the emission and absorption rates and the absorbed power
\begin{gather}
\label{Eq:GabsQuantum}
    \Gamma_{ab}=\frac{2\pi}{\hbar}\sum_{f,n}\frac{\omega_n}{2\epsilon_0V_n}d_{if}^2f(\epsilon_i)(1-f(\epsilon_f))n_{\rm{ph}}(\omega_n)\delta(\epsilon_f-\epsilon_i-\omega_n);\\
\label{Eq:GemQuantum}
    \Gamma_{em}=\frac{2\pi}{\hbar}\sum_{f,n}\frac{\omega_n}{2\epsilon_0V_n}d_{if}^2f(\epsilon_i)(1-f(\epsilon_f))(1+n_{\rm{ph}}(\omega_n))\delta(\epsilon_f-\epsilon_i+\omega_n);\\
\label{Eq:wQuantum}
    w=\frac{2\pi}{\hbar V}\sum_{i,f,n}\frac{\omega_n^2}{2\epsilon_0V_n}d_{if}^2f(\epsilon_i)(1-f(\epsilon_f))[n_{\rm{ph}}(\omega_n)\delta(\epsilon_f-\epsilon_i-\omega_n)-(1+n_{\rm{ph}}(\omega_n))\delta(\epsilon_f-\epsilon_i+\omega_n)].
\end{gather}
Exchanging $i\leftrightarrow f$ in the second term, one obtains
\begin{gather}
\notag    w=\frac{2\pi}{\hbar V}\sum_{i,f,n}\frac{\omega_n^2}{2\epsilon_0V_n}d_{if}^2\delta(\epsilon_f-\epsilon_i-\omega_n)[f(\epsilon_i)(1-f(\epsilon_f))(1+n_{\rm{ph}}(\omega_n))-(1-f(\epsilon_i))f(\epsilon_f)n_{\rm{ph}}(\omega_n)];\\
\label{Eq:wQuant}    w=\int d\varepsilon d\omega\sum_n\frac1{V_n}\delta(\omega-\omega_n)\frac1V\sum_{i,f}d_{if}^2\delta(\varepsilon-\epsilon_i)\delta(\epsilon_f-\varepsilon-\omega)
    \frac{\pi\omega^2}{\hbar\epsilon_0}[f_{\varepsilon}(1-f_{\varepsilon+\omega})(1+n_{\rm{ph}})-(1-f_{\varepsilon})f_{\varepsilon+\omega}n_{\rm{ph}}].
\end{gather}

If the electrons are in thermal equilibrium at temperature $T_e$, it is possible to write
\begin{gather}
\notag    f_{\varepsilon}(1-f_{\varepsilon+\omega})(1+n_{\rm{ph}}(\omega))-(1-f_{\varepsilon})f_{\varepsilon+\omega}n_{\rm{ph}}(\omega)
    =(f_{\varepsilon}-f_{\varepsilon+\omega})n_{\rm{ph}}(\omega)-(1-f_{\varepsilon})f_{\varepsilon+\omega};\\
\notag    f_{\epsilon}-f_{\epsilon+\omega}=\frac{e^{\frac{\varepsilon+\omega}{T_e}}-e^{\frac{\varepsilon}{T_e}}}{(e^{\frac{\varepsilon}{T_e}}+1)(e^{\frac{\varepsilon+\omega}{T_e}}+1)}=\frac{e^{\frac{\varepsilon}{T_e}}}{e^{\frac{\varepsilon}{T_e}}+1}\frac{1}{e^{\frac{\varepsilon+\omega}{T_e}}+1}(e^{\frac{\omega}{T_e}}-1);\\
\notag    (1-f_{\varepsilon})f_{\varepsilon+\omega}=\frac{e^{\frac{\varepsilon}{T_e}}}{e^{\frac{\varepsilon}{T_e}}+1}\frac{1}{e^{\frac{\varepsilon+\omega}{T_e}}+1}=\frac{f(\epsilon)-f(\epsilon+\omega)}{e^{\frac{\omega}{T_e}}-1}=(f(\epsilon)-f(\epsilon+\omega))n_B(\omega,T_e);\\
\label{Eq:f1f}    f_{\varepsilon}(1-f_{\varepsilon+\omega})(1+n_{\rm{ph}}(\omega))-(1-f_{\varepsilon})f_{\varepsilon+\omega}n_{\rm{ph}}(\omega)=[f(\varepsilon)-f(\varepsilon+\omega)][n_{\rm{ph}}(\omega)-n_B(\omega,T_e)],
\end{gather}
where $n_B(\omega,T_e)$ is the Bose-Einstein distribution for particles at energy $\omega$ and temperature $T_e$. Combining together Eqs. \eqref{Eq:sigma}, \eqref{Eq:wQuant} and \eqref{Eq:f1f}
\begin{gather}
\notag    w=\int d\varepsilon d\omega\sum_n\frac1{V_n}\delta(\omega-\omega_n)\frac1V\sum_{i,f}d_{if}^2\delta(\varepsilon-\epsilon_i)\delta(\epsilon_f-\varepsilon-\omega)
    \frac{\pi\omega^2}{\hbar\epsilon_0}[f(\varepsilon)-f(\varepsilon+\omega)][n_{\rm{ph}}(\omega)-n_B(\omega,T_e)];\\
\label{Eq:wQuant1}    w=\int d\omega\sum_n\frac1{V_n}\delta(\omega-\omega_n)\frac{\sigma(\omega,T_e)}{\epsilon_0}\omega[n_{\rm{ph}}(\omega)-n_B(\omega,T_e)],
\end{gather}
which corresponds to Eq. (1) of the main text with $D_{\rm{ph}}(\omega)\equiv\sum_n\frac1{V_n}\delta(\omega-\omega_n)$.

This is the heat power absorbed by the system per unit of volume due to the photon environment. When the background is thermal with $T_{\rm{ph}}=T_e$ the power vanishes, as it should since system and photons are in thermal equilibrium. The units are correct as $D_{\rm{ph}}$ is (energy$\times$volume)$^{-1}$, $\epsilon_0/\sigma$ has units of time and $\omega$ is an energy.

The treatment of matter systems made of bosons is equivalent, and only requires replacing $f_{\epsilon}\rightarrow n_B(\epsilon)$ and $1-f_{\epsilon}\rightarrow1+n_B(\epsilon)$. An equality equivalent to Eq. \eqref{Eq:f1f} can be derived for bosons, so that an equation similar to \eqref{Eq:wQuant1} holds with $\sigma$ representing the electric conductivity of the bosons. For a system of both fermions and bosons, it suffices to consider the total ac conductivity of the system.

\section{Calculation of the photon Density of States}\label{App:phDoS}

This section details the calculations for the photon DoS in the three cases used in the main text: \emph{(a)} free space; \emph{(b)} cavity with infinite planar mirrors; \emph{(c)} confined cavity with spherical mirrors. Working in SI units, the factors of $\hbar$ and $c$ are explicitly written.

\subsection{Free space}\label{App:phDoSfree}

This is the standard calculation of the photon density of states in the the black-body radiation derivation. The cavity volume goes to infinite, so it can be modelled as a cubic box with side $L$ so that $\omega_n=\frac{\pi\hbar c}L\sqrt{n_x^2+n_y^2+n_z^2}$ where $k_i=\frac{\pi n_i}{L}$ is the wavevector in the direction $i$ and $V_n=L^3$ for all modes. Thus
\begin{gather}
\notag D_{\rm{ph}}(\omega)=\sum_n\frac{1}{V_n}\delta(\omega-\omega_n)=\frac{2}{L^3}\sum_{n_x,n_y,n_z>0}\delta(\omega-\omega_n);\\
\label{Eq:Dfreespace}D_{\rm{ph}}(\omega)=\frac{2}{L^3}\frac{L^3}{\pi^3}\int_{k_x,k_y,k_z>0}d^3k\delta(\omega-\hbar c|\bf k|)=\frac{1}{4\pi^3}\int d^3k\delta(\omega-\hbar c|\bf k|)=\frac{1}{\pi^2}\int_0^{\infty}k^2dk\delta(\omega-\hbar ck)=\frac{\omega^2}{\pi^2\hbar^3c^3},
\end{gather}

where the factor two in the first line comes from the two photon polarizations.

\subsection{Infinite mirrors cavity}\label{App:phDoSplanar}

This subsection considers a cavity with planar mirrors at a distance $L$ between them and with surface area $A$. The volume is the same for all modes $V_n=AL$ and the mirrors extension in the transverse directions is much larger than their distance -- i.e. $A\gg L^2$ -- so that the sum over longitudinal modes is discrete, while the sum over the transverse modes can be approximated by an integral. The dispersion relation is $\omega_n(k_x,k_y)=\hbar c\sqrt{(n\pi/L)^2+k_x^2+k_y^2}$, where $n=1,2,3,...$ for TE modes and $n=0,1,2,...$ for TM modes. The fundamental cavity frequency is then $\omega_c=\hbar c\pi/L$
\begin{gather}
\notag D_{\rm{ph}}(\omega)=\sum_n\frac1{V_n}\delta(\omega-\omega_n)=\frac{1}{AL}\int \frac{A}{(2\pi)^2}dk_xdk_y\left(\delta(\omega-\hbar c\sqrt{k_x^2+k_y^2})+2\sum_{n>0}\delta(\omega-\omega_n(k_x,k_y))\right);\\
\notag D_{\rm{ph}}(\omega)=\frac{2\pi}{4\pi^2L}[\frac{\omega}{(\hbar c)^2}+2\int_0^{\infty} k_{\perp}dk_{\perp}\sum_n\delta(\omega-\hbar c\sqrt{\left(\frac{n\pi}L\right)^2+k_{\perp}^2})]=\frac{1}{2\pi(\hbar c)^2L}[\omega+\sum_n\int_{0}^{\infty} du\delta(\omega-\sqrt{(n\omega_c)^2+u})];\\
\label{Eq:D1D}D_{\rm{ph}}(\omega)=\frac{\omega_c}{2\pi^2(\hbar c)^3}\left(\omega+\sum_{n=1}^{\infty}2\omega\theta(\omega-n\omega_c)\right)=\frac{\omega\omega_c}{\pi^2\hbar^3c^3}\left(\left\lfloor\frac{\omega}{\omega_c}\right\rfloor+\frac12\right),
\end{gather}
where $u=(\hbar c k_{\perp})^2$, $\theta$ is the step function and $\lfloor\rfloor$ indicates the integer part.

For $\omega<\omega_c$, the phDoS of a planar cavity is linear in $\omega$, while that of free space is quadratic; however for $\omega>\omega_c$ they have the same order of magnitude. Note that for an infinite cavity, i.e. $L\rightarrow\infty$, $\omega_c\rightarrow0$ and the integer part is $\lfloor\omega/\omega_c\rfloor\approx\omega/\omega_c$ so that the phDoS reduces to the free space one.

In Eq. \eqref{Eq:D1D}, the integration over the transverse momentum $k_{\perp}$ extends from zero to infinity. This is strictly valid if the transverse direction of the cavity is infinite, otherwise there should be an upper cutoff because photons with large transverse momenta escape the cavity. This cutoff can be roughly estimated in the following way: consider a mode with longitudinal wavevector $n\omega_c/\hbar c$ and transverse wavevector $k_{\perp}$; the number of cycles that a photon lives inside the cavity is approximately the quality factor $Q$, so that the distance travelled in the longitudinal direction is $\sim QL$; thus the distance travelled in the transverse direction is $QLk_{\perp}/(n\omega_c/\hbar c)$; this distance should be smaller than the typical transverse dimensions of the cavity $\sim\sqrt{A}$: $\hbar c k_{\perp}\lesssim n\omega_c\sqrt{A}/QL$.

Thus the integral over $k_{\perp}$ is modified
\begin{gather*}
\int_0^{\infty} k_{\perp}dk_{\perp}\delta(\omega-\hbar c\sqrt{(n\pi/L)^2+k_{\perp}^2})\rightarrow\int_0^{n\omega_c\sqrt{A}/(\hbar cQL)} k_{\perp}dk_{\perp}\delta(\omega-\hbar c\sqrt{(n\pi/L)^2+k_{\perp}^2});\\
\theta(\omega-n\omega_c)\rightarrow\theta(\omega-n\omega_c)-\theta(\sqrt{(n\omega_c)^2+(n\omega_c)^2A/Q^2L^2}-\omega),
\end{gather*}
which means that $\omega$ must be comprised between $n\omega_c$ and $n\omega_c\sqrt{1+A/(QL)^2}$. Thus the phDoS is modified as
\begin{gather}
\label{Eq:D1Dperp}D_{\rm{ph}}(\omega)=\frac{\omega\omega_c}{\pi^2\hbar^3c^3}\left(\left\lfloor\frac{\omega}{\omega_c}\right\rfloor-\left\lfloor \frac{\omega}{\omega_c\sqrt{1+A/\left(QL\right)^2}}\right\rfloor\right).
\end{gather}

Note that the $1/2$ term arising from the TM$_0$ mode vanishes, since the TM$_0$ modes propagate parallel to the cavity mirrors and does do not stay inside the finite mirrors cavity. Except for this difference, which is significant only at small $\omega$, the qualitative behavior of the phDoS is not changed when accounting for the finite transverse size, as the large $\omega$ scaling is still $\sim\omega^2$, with an additional numerical prefactor smaller than one. However the small $\omega$ behavior of the phDoS becomes important only for large $\omega_c/k_BT$, which is not a relevant experimental regime for the scenarios we consider.For these reasons, the phDoS used in the main text is the one given by Eq. \eqref{Eq:D1D}.

\begin{figure}[t]
    \centering
    \includegraphics[width=0.96\columnwidth]{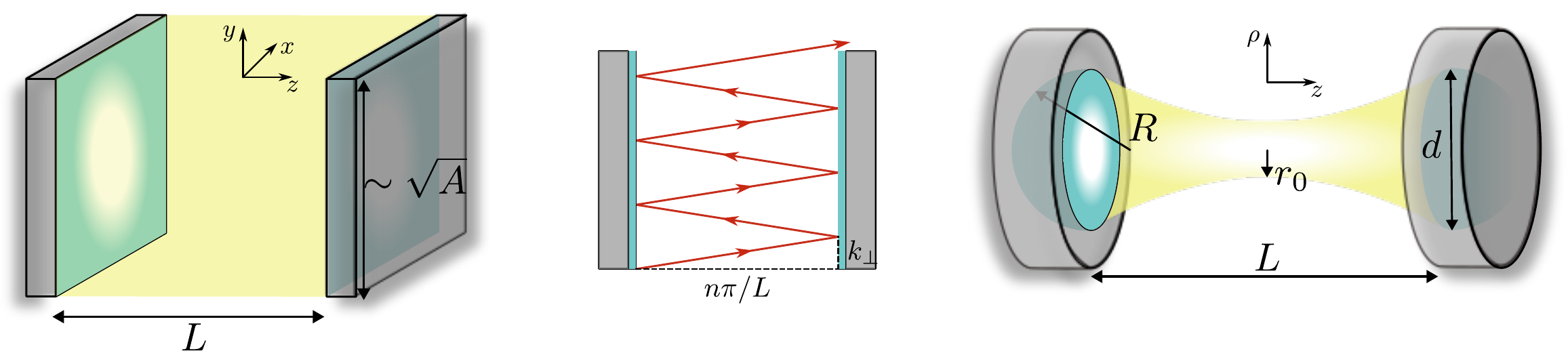}
\put(-480,96){(a)} 
\put(-320,96){(b)}
\put(-190,96){(c)}  
    \caption{(a) Sketch of a cavity with planar parallel mirrors, with length $L$ and transverse dimension $\sim\sqrt{A}$. (b) Schematic path of a photon mode with longitudinal wavevector $n\pi/L$ and transverse wavevector $k_{\perp}$. (c) Sketch of a confined cavity with spherical mirrors of diameter $d$ and curvature radius $R$, placed a distance $L$ apart. The beam radial size at the center of the cavity ($z=0$) is $r_0$.}
    \label{fig:CavitiesApp}
\end{figure}

\subsection{Confined cavity}\label{App:phDoSconc}

This Section deals with a confined cavity with two concave spherical mirrors of curvature radius $R$ situated at a distance $L$ along the $z$ axis. In the nearly concentric limit $R\approx L/2$. The modes $\rm{e}(\bf r)$ of the cavity are described in cartesian coordinates $(x,y,z)$ by a Hermite-Gaussian beam \cite{Photo:book,Utama2021:ConcHighTM}:
\begin{gather}
\label{Eq:LagGauss}
\rm e_{l,m}(\rho,\phi,z)=\frac{r_0}{r(z)}H_l\left(\frac{\sqrt2x}{r(z)}\right)H_m\left(\frac{\sqrt2y}{r(z)}\right)e^{-(x^2+y^2)^2/r(z)^2}
e^{i\Phi_{l,m}(x,y,z)}
e^{i\Phi_{l,m}(x,y,z)},
\end{gather}
where $H_l$ is the Hermite polynomial with index $l$. The width $r(z)$ describes the radial extension of the beam and assumes the value $r_0$ at the center of the cavity, $\Phi$ is a phase factor which determines the resonant frequencies. 

The modes frequencies are found to be
\begin{equation}
\label{Eq:ConcentricFreq}
\omega_{q,l,m}=\frac{\pi\hbar c}{L}\left(q+(l+m+1)\frac{\Delta\zeta}{\pi}\right)\approx\omega_c(q+l+m+1),
\end{equation}
where $q,l,m=0,1,2,...$, $q$ is the longitudinal quantum number and $\Delta\zeta$ is the Guoy phase shift inside the cavity; $\Delta\zeta$ approaches $\pi$ for nearly concentric cavities for which the beam waist at the center of $r_0$ goes to zero. In this regime the fundamental frequency of the cavity is $\omega_c=\pi\hbar c/L$.

The mode volume is found from the condition $\int d^3r|e_{q,l,m}^2|=V_{q,l,m}$. Operating a change of variable to $u=2\rho^2/r(z)^2$, it can be showed that for generic $p$ and $l$, this reduces to the orthogonality relations of the Hermite polynomials. The mode volume is
\begin{equation}
\label{Eq:ConcentricNorm}
V_n=V_{q,l,m}=\frac{\pi r_0^2(q,l,m)L}{2},
\end{equation}
where $r_0(q,l,m)$ for a possible dependence of the volume on the mode. Note that this volume is smaller than the physical volume of the cavity, because it is the effective volume occupied by the electromagnetic field.

Accounting for the two photon polarizations, the phDoS is written as
\begin{gather}
\label{Eq:DConcentric}
D_{\rm{ph}}(\omega)=\frac{4}{\pi r_0^2L}\sum_{q,l,m}\delta(\omega-\omega_c(1+q+l+m))=\frac{\omega_c\omega_{\rho}^2}{\pi^2\hbar^3c^3}\sum_{q,l,m}\delta(\omega-\omega_c(1+q+l+m)),
\end{gather}
where $\omega_{\rho}=2\hbar c/r_0$ is the energy scale associated to the beam waist in the radial direction. Equation \eqref{Eq:DConcentric} is valid when $\omega_{\rho}$ is constant; in general however, there is a hidden dependence on the modes, since the waist radius $r_0(q,l,m)$ it depends on the Rayleigh range and scales as $r_0\sim\sqrt{\lambda L}\sim1/\sqrt{\omega_{q,l,m}\omega_c}$. Writing $r_0(q,l,m)=\pi\sqrt{2\lambda L}/s$ -- where $s$ is a parameter depending on the exact geometry of the cavity -- so that $\omega_{\rho}^2=s^2\omega\omega_c$, Eq. \eqref{Eq:DConcentric} becomes
\begin{gather}
\label{Eq:DConcentric2}
D_{\rm{ph}}(\omega)=\frac{s^2\omega_c^2\omega}{\pi^2\hbar^3c^3}\sum_{q,l,m}\delta(\omega-\omega_c(1+q+l+m)),
\end{gather}

In the large $\omega\gg\omega_c$ limit, $\sum_{q,l,m}\rightarrow\int dqdldm$, and the phDoS becomes $D_{\rm{ph}}(\omega)\rightarrow\frac{\omega^2}{\pi^2\hbar^3c^3}\frac{\omega_{\rho}^2}{\omega_c^2}$, which differs from the free space phDoS by a factor $\omega_{\rho}^2/\omega_c^2$, or $D_{\rm{ph}}(\omega)\rightarrow\frac{s^2}{\pi^2\hbar^3c^3}\frac{\omega^3}{\omega_c}$ when accounting for the frequency dependence of $\omega_{\rho}$.

\section{Analysis of frequency dependent conductivity}\label{App:Condu}

In this section the behavior of Eq. (2) of the main text is investigated for two cases of frequency dependent conductivity:

\emph{(i)} a Drude metal with conductivity
\begin{equation}
\sigma_d(\omega)=\frac{\sigma_0}{1+\omega^2\tau_d^2}=\frac{\sigma_0}{1+\omega^2/\omega_d^2},
\end{equation}
where $\omega_d=1/\tau_d$ is the inverse of the elastic scattering time $\tau_d$.

\emph{(ii)} a system with a plasmonic-like optical resonance described by a Lorenz-Drude oscillator with resonance energy $\omega_0$ and damping $\gamma$. The conductivity is $\omega$ times the imaginary part of the response function:
\begin{equation}
\sigma_{\rm{pl}}(\omega)=\sigma_{0}\frac{\gamma\omega_{d}\omega^2}{[\gamma^2\omega^2+(\omega^2-\omega_0^2)^2]},
\end{equation}
where the presence of $\omega_d$ ensures that the conductivity satisfies the sum rule.

\begin{figure}[t]
    \centering
    {\includegraphics[width=\columnwidth]{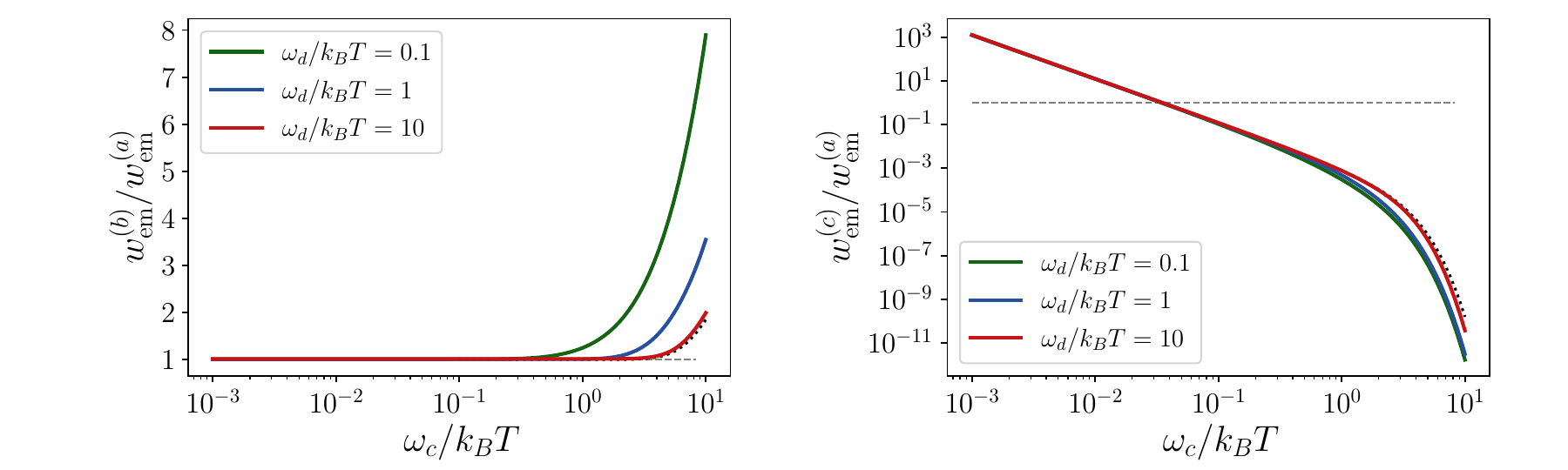}
    \put(-475,140){(a)} 
    \put(-245,140){(b)} }
    {\includegraphics[width=\columnwidth]{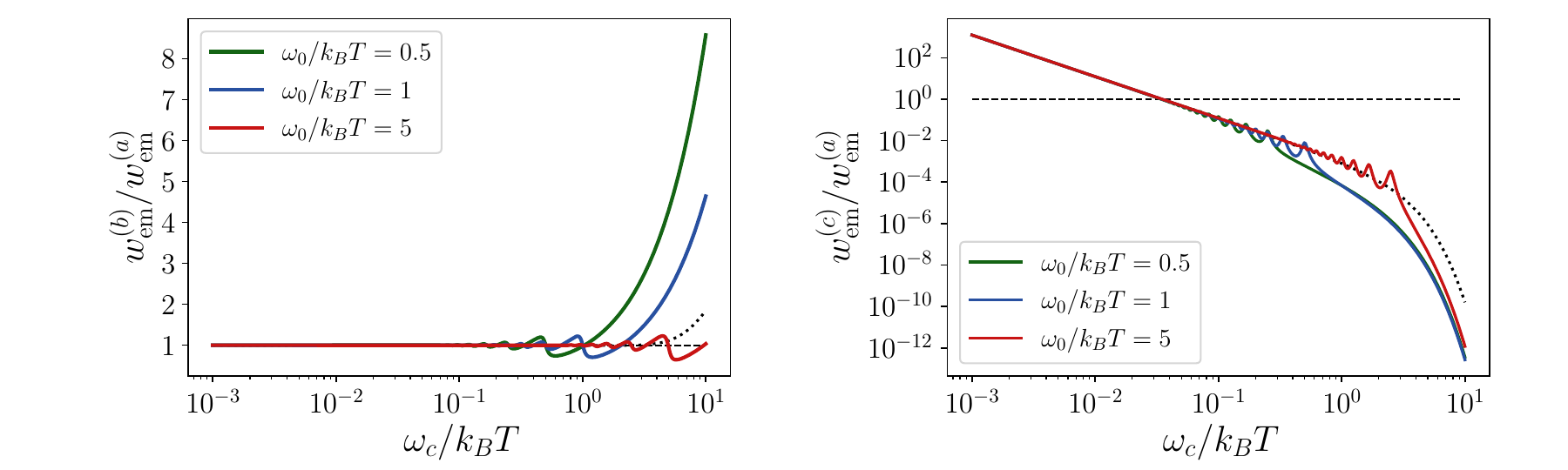}
\put(-475,140){(c)} 
\put(-245,140){(d)} }
    \caption{(a) Plot of the enhancement of the radiative power normalized to free space, as function of $\omega_c/k_BT$ for Drude conductivity and cavity with planar mirror, for different values of $\omega_d$. (b) Plot of the enhancement of the radiative power normalized to free space, as function of $\omega_c/k_BT$ for Drude conductivity and confined cavity, for different values of $\omega_d$ and $\omega_{\rho}/k_BT=0.05$. (c) Plot of the enhancement of the radiative power normalized to free space, as function of $\omega_c/k_BT$ for a plasmonic resonance conductivity and cavity with planar mirror, for different values of $\omega_0$. (d) Plot of the enhancement of the radiative power normalized to free space, as function of $\omega_c/k_BT$ for a plasmonic resonance conductivity and confined cavity, for different values of $\omega_0$ and $\omega_{\rho}/k_BT=0.05$. For both (c) and (d) the width is $\gamma=\omega_0/10$ and $\omega_d/k_BT=1$. The grey dashed line indicate the value of the free space power, while the dotted black lines are the values of the enhancement for a constant conductivity.}
    \label{fig:Conductivities}
\end{figure}

For the Drude metal case, the energy scale given by $\omega_d$ acts as a sort of temperature, so that when $\omega_d\gtrsim T$, there is no noticeable difference with the constant conductivity case analyzed in the main text. When $\omega_d\lesssim T$, it acts as a cut-off and the onset of the $\omega_c/T$ behavior of $\wem$ occurs around $\omega_d$. This is plotted in Fig. \ref{fig:Conductivities} a)-b). It can be seen that for $\omega_d/T=10$ there is no difference with the constant conductivity case.

For the case of the plasmonic-like resonance, the behavior of $\wem$ normalized to the free space case is plotted in Fig. \ref{fig:Conductivities} c)-d). The qualitative behavior is very similar to that calculated for a constant conductivity, except for the appearance of a series of peaks, which are the echoes of the resonance at $\omega_0$ in the conductivity and thus appear at $\omega_c=\omega_0/n$. The height and width of these peaks depends on $\gamma$. However the general trend of $\wem$ is not changed: for a cavity with planar mirrors $\wem$ is around one for $\omega_c\lesssim k_BT$ and increases with $\omega_c$ for $\omega_c\gtrsim k_BT$; for a confined cavity there is a decreasing power-law behavior at small $\omega_c$ and an exponential fall-off at $\omega_c\sim k_BT$, which occurs faster for smaller values of $\omega_0$.

These results confirm the statements made in the main text, i.e. that the essential phenomenology of $\wem(\omega_c,T)$ is captured by a constant conductivity and that more refined models for $\sigma(\omega)$ are only needed in the case of precise numerical estimates.

\end{document}